\documentstyle[epsf]{myl-aa} %for a non-referee version
\begin{document}

   \thesaurus{
%              (02.01.1;  % Acceleration of particles
%               09.03.2;  % cosmic rays
%               10.08.1   % Galaxy: halo
              ( 11.10.1;  % Galaxies: jets
               13.07.1)} % Gamma rays: bursts
\title { The cepheid-like relationship between variability and
luminosity explained within the ``cannonball
model'' of Gamma-Ray bursts }

   %\author{Arnon Dar\inst{1,2} and
   %        Rainer Plaga\inst{2}}
 \author{R. Plaga\inst{1}}

\offprints{plaga@mppmu.mpg.de}
  
\institute{
Max-Planck-Institut f\"ur Physik, F\"ohringer Ring
6, D-80805 M\"unchen, Germany }

   \date{Received, Accepted}
     
\maketitle
%\markboth{A.Dar \& R.Plaga:
%An alternative source of cosmic rays at all energies }
\markboth{Cepheid-like relationship within the cannon-ball model}
{}

\begin{abstract}
I show how an empirical variability - luminosity relationship
for prompt gamma-ray bursts,
first proposed by Fenimore and Ramirez-Ruiz, can be understood
as a special-relativistic beaming effect in the ``cannonball model''
of Dar and De R\'ujula. In this scenario the variability is a measure of the
direction of propagation and the Lorentz factor of the cannonball
on which in turn the apparent luminosity of the prompt GRB depends
sensitively.
The observed absence of cosmological time dilation
in the ``aligned peak test'' - when using redshifts derived with this
relation - is also explained. 
The most direct evidence in favour of the cannonball model
presented here is its correct description for
the observed relation between narrow-spike width and amplitude within a given 
GRB.
There seems to be an indication for cosmological time dilation in the
total duration of GRBs, as expected in the cannonball model.
Quantitative predictions for the luminosity function of GRBs and the
``spectral-lag luminosity relation'' are given.
\end{abstract}
%
%________________________________________________________________ 
\section{Introduction}
\label{intro}
\subsection{ Fireball, jet, cannonball: the evolutionary sequence of 
models for the basic geometry of a GRBs? }
\label{model}
\begin{figure}[ht]
\vspace{0cm}
\hspace{0cm} \epsfxsize=5.1cm 
\epsfbox{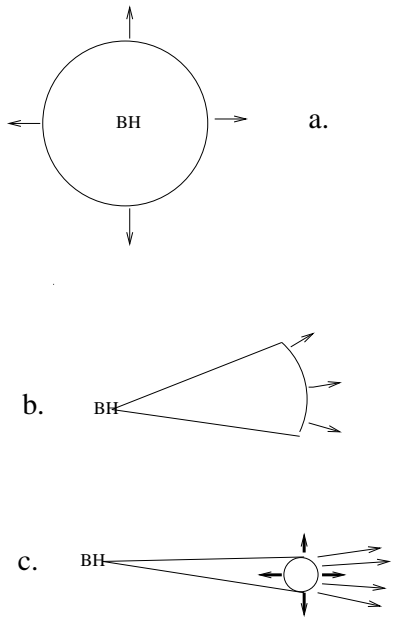}
\vspace{0cm}
\caption{Three options for the basic geometry of a GRB,
a. fireball/firecone: opening angle $\theta_o \approx \pi$ 
b. jet: $\theta_o \ll 1$ c. cannon ball: $\theta_o
\approx 1/(\Gamma \sqrt{3}$). ``BH'' stands
for the location of the compact object (most likely a Black Hole)
that ejects relativistic matter towards the observer.
The thin arrowed lines indicate the direction of the emitted radiation.
In a. and b. this is identical with the direction of motion of
the ejected matter to good approximation. The thick arrowed lines indicate
the direction of motion of matter in the cannon-ball´s rest frame.} 
\label{fig1}
\end{figure}

Gamma-ray bursts (GRBs) are short (durations down to less than a second)
and bright (fluxes up to 10$^{-4}$ erg/cm$^2$/sec, brighter than
any stationary source)
bursts of $\gamma$-rays
from distant galaxies that are detected by space satellites at a rate of
$R_{\rm GRB}$ $\sim 10^3$ per year (\cite{sngrb}).  
As was realised in two seminal papers (\cite{pac,good}),
at cosmological source distances,  
basic radiation physics dictates that the matter
emitting $\gamma$ radiation with these
durations and intensities moves at highly relativistic
speeds (Lorentz factors $\Gamma$ $>$ 300 (\cite{gammalim})).
Were this matter emitted isotropically (thus forming a ``fireball'')
the total $\gamma$-ray luminosity of some bursts
would be $>$ 10$^{53}$ ergs/sec -
a considerable fraction of the total gravitational 
energy of any stellar mass object. The problem of radiating
that much energy as MeV-$\gamma$-rays was already clearly
recognised as difficult to solve
by Paczy\'nski and Goodman (\cite{pac,good}) (neither did later authors
find a conversion mechanism efficient enough for this feat).
This dilemma even led Goodman to speculate 
that some ``exotic and unimagined process''
- rather than some type of stellar collapse - might power
the fireball.
\\
From 1992 onwards it began to be recognised 
(\cite{brainerd,mesjet,shav}) that
the relativistic outflow might not be isotropic but jet-like,
thus reducing the energy output of GRBs to a level more readily
producible in stellar collapse.
The total luminosity 
of blazars - active galaxies with a relativistic
jet pointed in our direction - 
{\it appeared} to be dominated by $\gamma$-rays until
it was realised that the total energy emitted in the form
of $\gamma$-rays is reduced
by a factor ``4$\pi$/(solid angle into which the jet emits the
observed $\gamma$-rays)''. 
Together with the close semblance of spectra
and time variability in GRBs and blazars this
suggested that GRBs are due
to jets; the properties of the afterglows of GRBs later made
this a virtual certainty for the subclass of ``long'' GRBs (\cite{dar98}).
A jet can be understood as an angular section of a fireball with
an opening angle $\theta_o$ $\ll$ 1. The basic physics of
a magnetically driven outflow from a collapsed stellar object
which would lead to a jet with $\theta$ $\approx$ 0.1
were discussed by Meszaros and Rees (\cite{rees96}).
Pugliese, Falcke \& Biermann (1999) discussed the detailed physical
mechanism that gives rise to GRBs in a jet powered by an accretion disk. 
Aloy et al.(1999) performed detailed numerical calculations of the formation
of ultrarelativistic jets in stellar collapse.
\\
Jets formed by black holes in the Galaxy and the centres of active galaxies
are observed to have a very
inhomogenous structure consisting of sharply defined 
irregular features that have been interpreted as
balls of plasma (for respective reviews see (\cite{mirabel,narl})).  
%This structure plausibly interpreted
%as being due to a black hole that accretes matter episodically with the subsequent
%emission of a ball of plasma (\cite{mirabel}).
Following Shaviv and Dar (1995) it was proposed that 
jets formed by gamma-ray bursters
have a similar structure. The GRBs 
are then emitted by ``blops'' (\cite{blackman}),
``plasmoids'' (\cite{chiang,darplaga}) or ``bullets'' (\cite{umeda})
that propagate with a Lorentz factor $\Gamma$ $\gg$ 1 approximately
along a single trajectory (i.e. they are emitted along
directions differing by angles $\ll$ 1/$\Gamma$).
\\
Recently Dar and De R\'ujula (2000a,2000b) 
developed a quantitative ``cannonball
model'' which builds upon this idea.
In this model episodic accretion onto a central
black hole (\cite{woosley93}) or neutron star leads to the subsequent 
repeated emission of 
distinct masses of plasma,the ``cannonballs'' 
initially moving with a Lorentz $\Gamma$ $\approx$ 1000.
Radiation emitted by the cannonballs gives rise to GRBs.
This radiation is emitted isotropically in the rest frame of
the cannonball, but in the observer frame the Lorentz boost
strongly collimates the radiation into a cone 
with opening angle 1/$\Gamma$
along the direction of motion (``relativistic beaming''). 
The connection between the sudden disappearance of matter through the
horizon of a black hole and the 
subsequent ejection of expanding clouds of
relativistic plasma - which forms the 
conceptual basis of the cannonball model -
was observationally studied in detail 
in the Galactic microquasar GRS 1915+105 (\cite{mirabel}).
Cannonballs expand in their rest frame
with the speed of sound in a relativistic gas
c/$\sqrt{3}$.
Kinematically ejected
cannonballs are equivalent to a non-stationary jet with 
a very small opening angle $\theta_o$=1/($\Gamma$ $\sqrt{3}$).
The model's decisive progress is that $\theta_o$ is now fixed by basic
physics rather than guesswork.
Because $\theta_o$ is smaller than the special-relativistic
beaming angle $\theta_r$=1/$\Gamma$, the 
emission anisotropy of ``cannonball
GRBs'' is {\it dominated at all viewing angles} by special-relativistic
effects.
This leaves no free parameters for its calculation,
except $\Gamma$ that is experimentally constrained from below.
The ``evolution''of GRB models is graphically illustrated in
fig. \ref{fig1}.

The following ``rate argument'' is
in favour of the basic cannonball geometry.
In the cannonball model
the {\it vast} majority of GRBs are beamed away from us, only a fraction
of order
$\theta^2$/4 $\approx$ 2 $\times$ 10$^{-7}$ ($\Gamma$/1000)$^{-2}$
is visible.
The observed rate of GRBs then corresponds
to a rate GRBs that is of
the same order of magnitude as the one of stellar collapse (\cite{darplaga}).
The emission of a GRB might then occur 
in {\it ``common''} supernovae (i.e. 
ratio (SNe with GRB emission)/(all SNe) is {\it not} $\ll$ 1)
(\cite{cen,wang98})
- rather than only
in very rare exceptional collapse events (\cite{pac97}). 
This conclusion found experimental support 
in the observation of several GRB-supernova 
associations in 1998/99 (\cite{sngrb})
and the possible observation of an ultra-relativistic jet
in the only Galactic SN of the past century: SN87A (\cite{pirannak,cenjet}).
Previously
possible associations of ``common-supernova'' with GRBs were suggested
by Colgate (1968) and motivated the search
for GRBs which led to their discovery in 1973 (\cite{disc}). 
The physics of non-axisymmetric type-Ib supernova explosions 
of Wolf Rayet stars had been discussed
by Biermann (1993) and Woosley (1993).
%The GRB - ``jet of common-supernova''
%connection was suggested
%after the experimental discovery of the association
%of GRB 980425 with SN1998bw (\cite{cen,wang98}).
\subsection{ The cepheid-like relation between variability and
luminosity in GRBs}
\label{cepheid}
Based on seven GRBs with optically determined afterglow-redshift
Fenimore and Ramirez-Ruiz (2000) found an empirical
relationship between the variability V in the prompt GRB
and its peak luminosity per steradian L/$d\Omega$ of
\begin{equation}
L/d\Omega = 3.1 \times 10^{56} V^{\delta} {\rm erg/sec}
\label{ceph}
\end{equation}
The preferred fit is obtained with $\delta$=3.35.
They defined variability ``V'' as the ``spikiness'' of 
gamma-ray light curve
in the prompt burst, i.e. the intensity of a high-frequency noise
commonly discernible in GRB time histories.
Mathematically V is defined as the normalised random
mean-square (RMS) of the GRB intensity I as a function of time,
after removing low frequencies by smoothing. 
V as defined by Fenimore and Ramirez-Ruiz (2000) can be written 
schematically as:
\begin{equation}
V \sim {\rm RMS}(I_{\rm f>f_S})/I_{\rm max}^2
\label{var}
\end{equation}
Here I$_{\rm max}$ is the peak intensity of the GRB time
history.
$f$ is frequency in the Fourier decomposition of a GRB time
history. ``High'' frequencies are defined to be the ones greater than $f_S$ $\approx$ 0.2 Hz.
Fenimore and Ramirez-Ruiz (2000) do not fix $f_S$,
but coose it inversly proportional to T$_{90}$ - the period that 
contains 90 $\%$ of all counts in a GRB time history.
Because the interburst 
variation in T$_{90}$ is not quantitatively
understood in the cannonball model, 
I can only consider a simplified case of
fixed T$_{90}$ and consequently fixed $f_S$ below.
\\
Another theoretical collaboration suggested a similar
relationship in a sample of 20 bursts for which direct information on the
distance is available
(\cite{reich}).
They specify the exponent of V in a relation similar
to eq (\ref{ceph}) as: 
\begin{equation}
\delta = 3.3^{+1.1}_{-0.6}
\label{delta}
\end{equation}
Fenimore and Ramirez-Ruiz (2000) 
went on to use the relationship to derive
redshifts for 220 long bright GRBs observed by BATSE,
finding values larger than z=12 and peak 
``isotropic'' peak luminosities $L/d\Omega$ $\times$ 4$\pi$ surpassing
10$^{55}$ erg/sec.
%If GRBs are linked to stellar collapse one expects that their
%rate closely follows the star formation rate.
%Remarkably Fenimore and Ramirez-Ruiz find that within the errors
%and below z=2 the rate of GRBs R - inferred assuming the
%``variability inferred'' redshifts - has the same functional
%dependence on z (R $\sim$ (1+z)$^{3.3}$)
%than the one inferred from completely independent
%observations on star formation.
\\
If confirmed, this cepheid-like relation promises to become
a powerful tool for cosmology because GRBs
- then assuming the role of standard candles -
can be observed with a high rate out to higher
redshifts than SN Ia. Moreover 
nowhere in the electromagnetic spectrum
are standard candles more valuable than at MeV energies
because nowhere is the universe more transparent.
However, Fenimore and Ramirez-Ruiz (2000) find that narrow
spikes in the
GRB time history are not stretched via time dilation
with rising z - as expected if variability
provides valid redshifts. This 
would seem to cast a major doubt on the reality
of the proposed relation eq.(\ref{ceph}).
\subsection{Aim of this paper}
\label{aim}

In this paper I show that the proposed relationship between
time variability and luminosity of GRBs eq.(\ref{ceph})
can be quantitatively understood
as a special-relativistic beaming effect within the cannonball
model. 
Moreover the main objection against the reality
of the proposed variability-luminosity relationship raised
by its authors
- the absence of time dilation 
mentioned in the previous section -
can be readily understood.

\section{ Variability - Luminosity relationship as a 
simple special-relativistic effect }

\subsection{Two mechanisms of GRB variablity in the cannonball 
model: episodic accretion (at all frequencies) and instabilities in the CB
(at high frequencies)}
\label{varphys}
\begin{figure}[ht]
\vspace{0cm}
\hspace{0cm} \epsfxsize=8.1cm 
\epsfbox{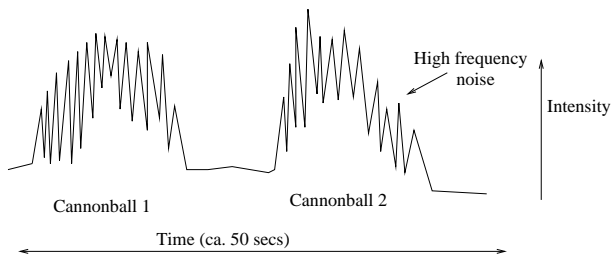}
\vspace{0cm}
\caption{Cartoon of the time history of a GRB
created by the ejection of 2 cannonballs.
The two broad ``subpulses'' originate in the ejection of
two balls of plasma from a central black hole. 
Their distance in time is
stretched only by cosmological time dilation $\sim$ (1+z).
I argue in the text that observational data indicate that
the high-frequency noise, consisting
of short ``spikes'' with typically ca. 1 sec FWHM, is due to
turbulent inhomogeneities in the moving cannonballs.
The intervals between these spikes and their
widths are then in addition proportional to the special
relativistic factor F (eq.(\ref{F})).} 
\label{fig2}
\end{figure}

In the cannonball model the observed variability in 
the prompt GRB can be due to two mechanisms, illustrated 
in fig. \ref{fig2}.
\\
First the ejection of distinct
cannonballs leads to distinct broad ``subpulses'' in the GRB
time history.
At low frequencies ($f$ $<$ $f_S$ $\approx$ 0.2 Hz) variability is
dominated by this mechanism.
\\
Second it seems possible
that at high frequencies  ($f$ $>$ 0.2 Hz)
magnetohydrodynamic instabilities 
in the {\it moving} cannonball contribute (\cite{shav}).
Observational arguments for the development of turbulence
in the cannonballs is discussed in the next section.
This mechanism leads to narrow ``spikes'' in the time history.
\\
Both mechanisms are affected by cosmological time dilation.
In addition the second mechanism (but not the first one)
is affected by special relativistic
effects. 
The experimental data discussed in section \ref{intra}
make it likely that at high frequencies the second mechanism
actually {\it dominates}.
%The power spectrum of intensity variations in radio jets from
%active galaxies - where such instabilites have been frequently discussed -
%is found to follow a power-law with an index of -2 very
%similar to the one found in GRBs (\cite{shav}).
\subsection{Variability 
quantitified: the power density spectrum of GRBs as evidence
for turbulence}
\label{power} 
The power density spectrum
of the prompt ``long'' GRB time histories
as a function of frequency is found to be
described by a power law 
\begin{equation}
P \sim f^d 
\label{powerspec}
\end{equation}
to good approximation 
between ca. 0.03 and 1 Hz (\cite{shav,belli}).
Shaviv and Dar (1995) found a typical index d = $-$2  
in BATSE data, while 
Beloborodov Stern $\&$ Svensson (2000) find indices ranging from
$-$1.5 to $-$1.82 for different intensity and energy groupings.
The exact index seems to depend on details of the burst selection
and analysis method.
At higher and lower frequencies the spectrum breaks off.
Such a power law is expected to arise in a turbulent process;
the expected power-law indices for Kolmogoroff and Kraichnan
turbulence are $-$1.67 and $-$1.5 respectively, the one for Brownian motion
is $-$2.

%Variability as defined by Fenimore and Ramirez-Ruiz (``spikiness'')
%measures the power at high frequencies and is thus expected to be 
%determined by processes in the moving cannoball.

\subsection{Time - intensity structure of
``spikes'' in the cannonball model}

In the previous section I argued that
narrow spikes in the prompt GRB light curve
can in principle be due to inhomogeneities in a cannonball
that moves with a speed corresponding to a Lorentz factor
$\Gamma$ $\gg$ 1.
The angle $\theta_v$ $\ll$ 1 between the direction of motion of the
inhomogeneity and the observer then influences
both temporal width (duration) W and intensity of the spike I$_s$
via the following well known special-relativistic relations 
(see (\cite{narl}) for a review of bulk relativistic motion).
All time scales corresponding to Fourier frequencies $>$ $f_S$,
including the widths of the spikes W, are compressed
proportional to F:
\begin{equation}
W \sim F
\label{width}
\end{equation}
All intensities, including the ones at the spike I$_s$, are boosted
according to: 
\begin{equation}
I_s \sim F^{-3-\alpha}
\label{intensity}
\end{equation}
with 
\begin{equation}
F = (1 + \theta_v^2 \Gamma^2)/2\Gamma
\label{F}
\end{equation}
and $\alpha$ the power-law index of the intensity
spectrum.
I always set $\alpha$ = 0 below, this is correct for the integral
of intensity over electromagnetic-radiation frequency $\nu$. 
The  latter condition is approximately fulfilled: 
the BATSE DISSC data - used
by Fenimore and Ramirez-Ruiz (2000) - 
are for a large spectral interval (25 - 800 keV).
$\theta_v$ is {\it not} expected to be constant within a 
given GRB in the cannonball model because
the ``edges'' of the cannonball 
subtend an angle
of 2/($\Gamma$ $\sqrt{3}$) with the burst centre. 
If the variation in spike durations in a given GRB would be {\it dominated} 
by beaming effects we expect spike widths varying by 
up to about a factor 2 for $\theta_v$, not very different from
1/$\Gamma$. Such small $\theta_i$ mainly occur because
they are strongly Doppler favoured
via eq (\ref{intensity}).
Combining eq.(\ref{width}) and eq.(\ref{intensity}) 
the spike intensity W should 
depend on I via:
\begin{equation}
I_s \sim W^{-3}
\label{iw}
\end{equation}

\subsection{Intraburst variability: 
``spike'' amplitude as a function of width}
\label{intra}
Ramirez-Ruiz and Fenimore (1999) 
have recently studied the relationship between
short-spike amplitude and width within a single GRB time histories. 
In a sample
of 387 spikes in 28 bright ``long'' GRBs they find:
\begin{equation}
I_s/<I_s> \sim [W_{FWHM}]^{-2.8}
\end{equation}
for $W_{FWHM}$ from 0.6 to 1.3 seconds.
The exponent is $-$3.0 with a slightly different fitting
procedure.
This result is in complete agreement with the prediction
of eq.(\ref{iw}) both in functional dependence and range of
W.
In my mind the correct explanation
of the intra-burst width - intensity relation is the strongest
evidence in favour of the cannonball model discussed in this paper.
\\
Spike widths around a second contribute
to high frequencies around 1 Hz.
The agreement is thus evidence
that at high frequencies 
the variability is {\it dominately} 
caused within the moving cannonball and is
thus subject to special-relativistic beaming effects.

\subsection{Interburst variability: Apparent luminosity as a function
of variability}

In order to predict the
relation between peak intensity 
I$_{max}$ and variability V of a GRB in the cannonball
model, we must infer their respective dependence on F (eq.\ref{F}).
% int = f(F)
Two effects are important:
\\
1. Relativistic beaming compresses the GRB time history at frequencies
above $f_S$ proportional to F, according to equation (\ref{width}).
\\
2. All intensities are boosted according to equation (\ref{intensity})
\\
I$_{max}$ is not influenced by effect ``1'' because the 
broad ``subpulse''
peak intensity, which dominates the GRB peak intensity, corresponds
to a frequency $f$ $<$ $f_S$.
Effect ``2'' alone 
then leads to the following dependence in the observer frame:
\begin{equation}
I_{\rm max} \sim F^{-3}
\label{inttot}
\end{equation}
% V = f(F)
If the phases of the Fourier components of the GRB time history
are uncorrelated, 
$RMS(I_{\rm f > f_S})$ is given as the quadratic sum of the
Fourier components with $f$ $>$ $f_S$.
In the continuous case the following proportionality then holds
in the cannonball restframe:
\begin{equation}
 RMS(I_{\rm f > f_S })_{\rm rest} 
\sim \int^{\infty}_{f_S} P(f) df 
\label{hipo}
\end{equation}
When we go to the observer frame
effect ``1'' stretches all frequencies above $f_S$ by a 
a factor F$^{-1}$, i.e. a power density spectrum P(f/F) 
instead of P(f) has to be used to evaluate
$RMS(I_{\rm f > f_S })_{\rm observer}$, the RMS of the high
frequency noise in the external observer frame.
For a power law only the normalisation 
is changed by this transformation. 
Effect ``2'' increases the intensities at all Fourier
frequencies by a factor F$^{-3}$,
so that the right-hand side of eq. (\ref{hipo}) 
has to be multiplied by a factor
F$^{-6}$. One finally obtains:
\begin{equation}
 RMS(I_{\rm f > f_S })_{\rm observer} 
\sim \int^{\infty}_{f_S} P(f/F) F^{-6} df 
\label{hipo2}
\end{equation}
Replacing P by its power-law expression eq.(\ref{powerspec})
and inserting eq.s (\ref{inttot}) and
(\ref{hipo2}) into eq.(\ref{var}) yields:
\begin{equation}
V \sim  
\int^{\infty}_{f_S} (f/F)^d df \sim 
(F/f_S)^{(-d-1)}
\label{var2}
\end{equation}
The quantity V, as defined by Fenimore and
Ramirez-Ruiz (2000), depends on F in my scenario.
Because the intensity also depends on F
we can combine eq.(\ref{var2}) with eq.(\ref{inttot}) and get:
\begin{equation}
I_{\rm max} \sim V^{3\over(-d-1)}
\label{cephtheo}
\end{equation} 
For d=$-$2 this results in 
I$_{\rm max}$ $\sim$ V$^3$ for d=$-$5/3: I$_{\rm max}$ $\sim$ V$^{4.5}$.
Thus - for values of d bracketing the plausible range of d 
(see section \ref{power})-
the resulting expression explains the empirical
variability-luminosity relationship eq (\ref{ceph}) 
within the errors (given in relation (\ref{delta})). The 
cepheid relation
is then understood if the luminosity of all cannonballs
at $\theta_v$=0 is constant among different GRBs.
Dar and De R\'ujula (2000a) argue that this indeed holds true
to some degree.
The data points show a great scatter with respect to
eq. (\ref{ceph}), this may be partly due to deviations
from this ``standard-candle'' assumption.

\subsection{How a luminosity bias cancels the expected redshift effect
in the ``aligned pulse'' test}
\label{lumicanc}
Fenimore and Ramirez-Ruiz (2000) tested their proposed
variability - luminosity relationship by searching
for a cosmological time dilation effect on the 
spike pulse widths W in GRBs. After correcting
for an empirical energy dependence of pulse width
on energy one expects the following
cosmological dependence of mean spike width W on redshift:
\begin{equation}
W \sim (1 + z_{\rm vi})^{-0.42}
\label{stretch}
\end{equation}
Here z$_{\rm vi}$ is the redshift inferred from the 
variability of the burst. This should result in a sensitive
test for the redshifts up to 7 that were used.
\\
This expected stretching with redshift was not found in the data.
What was found was
even some evidence for a small ``anti-effect'' i.e.
the peaks seem to get slightly narrower with rising redshift!
The prediction of the cannonball model is different from
eq. (\ref{stretch}).
One has to take into account that
the bursts with large z tend to have a higher luminosity
and thus smaller F. This bias towards
small F narrows the pulses
thus counteracting the cosmological ``stretch effect''.
Quantitatively I find for the mean peak luminosity
of the first redshift interval used by them
(0.3 $<$ z$_{vi}$ $<$ 0.75)
0.4 $\times$ 10$^{52}$ erg and for the last one (5 $<$ z$_{vi}$ $<$ 7)
7.0 $\times$ 10$^{52}$ erg. This translates into an ``anti-stretch'' factor
of (7.0/0.4)$^{1/3}$ = 2.6 via eq (\ref{iw}). This more than offsets
the expected cosmological stretch of (7/1.5)$^{0.58}$=2.4 among
burst in the mentioned z$_{vi}$-intervals. 
Taking into account the luminosity bias, instead of eq.(\ref{stretch})
the cannonball
model thus predicts a combined
general- and special-relativistic z - dependence of the 
short spike width W in the ``aligned
peak test'' of: 
\begin{equation}
W \sim (1 + z)^{-0.1}
\label{stretch2}
\end{equation} 
Spikes of bursts in 
``high redshift'' group are expected to be about 15 $\%$ {\it narrower} 
than the ones in the ``low redshift'' group.
This in excellent agreement
with the ``aligned peak test'' reported by Fenimore and 
Ramirez - Ruiz (2000) fig. 11.

\section{ Unsettled issues and predictions }

\subsection{ The luminosity function of GRBs }

Only relativistic beaming seems capable
of explaining the huge spread in observed GRB-luminosities (6.3 (2.5)
orders of magnitude with (without) GRB 980425 in the
small group of GRBs with directly measured redshifts) which do
not seem to {\it dramatically} influence other burst properties.
\\
What is more, relativistic beaming 
of ``standard-candle cannonballs''
makes a quantitative prediction for the luminosity
function.
It is well known that relativistic beaming of a point source
produces a luminosity function $\phi$ described 
by a power law without breaks (e.g.\cite{yi}):
\begin{equation}
\phi(L/L_0) \sim (L/L_0)^{-5/4}
\label{lf}
\end{equation}
Fenimore and Ramirez-Ruiz (2000)
deduced the observed 
luminosity function of GRBs using variability
inferred luminosities and find that it is described by a power law.
They argue that ``we might be seeing a
gradual roll over ... at low luminosities'', but I can see
no evidence for this in their figure 10.
Performing an unbroken power-law fit to their star-formation (GRB formation)
corrected luminosity function
over the whole luminosity range I find power law indices
of  of $-$1.5 ($-$1.9). One has to await error estimates
for the luminosity function of Fenimore and Ramirez-Ruiz
to quantitatively compare these values with the
predicted in eq. (\ref{lf}) (i.e. $-$1.25). However considering
the possible sources of systematic errors (
uncertainties in: cosmological model, variability-luminosity
relation, corrections for detection efficiencies and spectra
of GRBs) I find the present agreement satisfactory.

\subsection{ Direct evidence for cosmological redshifts in the
T$_{90}$ GRB durations? }

From the discussion in section \ref{varphys} we expect
that the cosmological ``time dilation'' effect - though
offset by special relativistic effects at high frequencies 
(section \ref{lumicanc}) -
should be measurable at low frequencies, where the duration
is determined by episodic accretion.
Because this process takes place in the same inertial frame
as the observer to good approximation, it cannot be influenced
by special-relativistic effects.
In particular T$_{90}$ -  the period that 
contains 90 $\%$ of all counts - should be
cosmologically time-dilation stretched:
\begin{equation}
T_{90} \sim 1+z
\label{strez}
\end{equation}
There is no significant dependance of the hardness ratio on
T$_{90}$ (\cite{kouve}), so here we expect no large energy dependent
corrections.
\\
In table \ref{table1} (extracted
from table 1 of (\cite{fenimore2000})) the mean z and T$_{90}$ for the
two even sized GRB samples with directly
measured redshift (with z $<$ 0.96 and z $>$ 0.96) are reported.
A major difficulty for studies using T$_{90}$ is a remaining ``tip of the
iceberg'' effect, the fact that dim GRBs have a bias towards
short T$_{90}$ values because portions of the burst drown 
in the background. Therefore bursts with a ``peak photon-flux/cm$^2$/sec 
on the 256 msec
time scale'' P$_{256}$ $>$ 10 (i.e. GRB 990123) were excluded from the
sample. 
\\
Clearly the T$_{90}$ values have a large scatter and are not
only dependent on 1+z.
However, the ratio of the mean of the 
T$_{90}$ values of the two groups
T$_{90}$(z $>$ 0.96)/T$_{90}$(z $<$ 0.96) = 3.4 $\pm$ 1.5
is in agreement with the ratio of mean 1+z values 
mean[1+z](z $>$ 0.96)/mean[1+z](z $<$ 0.96)=1.9.
The 
difference of the mean T$_{90}$ values (33.5 seconds) is different
from 0 on the 3.4 $\sigma$ level. 
Taking into account the possibility of systematic errors
of this very rough analysis, this evidence is 
inconclusive but suggestive.
\begin{table}[h]
\vspace{-10pt}
\caption{Redshift z,  duration T$_{90}$ (in seconds) and peak rate P$_{256}$
(in photons/cm$^2$/sec)
for 6 GRBs with directly determined redshifts with the mean values
for two redshifts groups.}
\label{table1}
\begin{center}
\begin{tabular}{lcccc}
Group with z $<$ 0.96 & & & &\\
\hline\hline
GRB date & z & T$_{90}$ &  P$_{256}$ \\
\hline
980524 &  0.0085 & 20.6  & 1.1   \\
970508 &  0.835 &  13.2  & 1.2   \\
970828 &  0.958 &  8.8   & 4.9   \\
  \\
\hline
mean & 0.60 & 14.2 $\pm$ 3.5 & 2.4 $\pm$ 1.2 \\ 
\hline
   \\
Group with z $>$ 0.96 & & & &\\
\hline\hline
GRB date & z & T$_{90}$ &  P$_{256}$ \\
\hline
989793 &  0.967 & 51.2  & 2.6   \\
990510 &  1.62 &  60.6  & 8.2   \\
971214 &  3.412 & 29.9   & 2.3   \\
  \\
\hline
mean & 2.0 & 47.7 $\pm$ 9.2 & 4.3 $\pm$ 1.9 \\ 
\hline
\end{tabular}
\end{center}
\end{table}
\\
Using ``variability-derived'' redshifts 
(\cite{fenimore2000})
to look for cosmological time dilation in T$_{90}$ is 
a procedure even more prone
to systematic errors because the GRB sample used by them was
selected using cuts on T$_{90}$, recorded-pulse quality and P$_{256}$
(the sample of ``GRBs with directly determined redshifts'' is also
not completely free of such biases e.g. via the T$_{90}$ dependent trigger
efficiency of Beppo-SAX).
%However, it is still worthwhile to point out that a simple
%analysis gives a very significant effect of the expected magnitude.
%I selected two samples of 10 GRBs
%with the same cuts on z and P$_{256}$. In addition a cut
%of z $<$ 3.5 (the highest directly determined redshift).
%The results are reported in table ($\ref{table2}$).
%T$_{90}$(z $>$ 0.96)/T$_{90}$(z $<$ 0.96) = 2.2 $\pm$ 1.1
%to be compared with an expected value of  
%mean(1+z)(z $>$ 0.96)/mean(1+z)(z $<$ 0.96)=1.7.
%These numbers are in acceptable agreement and here the
%difference in mean T$_{90}$ (46 sec) is significant on
%the 5 $\sigma$ level.
%The mean T$_{90}$ does not increase for samples with 
%higher z in the expected amount, rather it remains fixed
%at about 90 seconds.
%This could signal a breakdown of the variability - luminosity
%relation in its present form for z $>$ 3.5 or a selection effect
%due to a reduction of the mean P$_{256}$ by the cosmological stretching.
\\
The issue of cosmological redshifts affecting T$_{90}$
needs more work. If a careful analysis finds evidence
for time dilation in this quantity with ``variability - derived''
redshifts, this might allow
to calibrate the variability - luminosity relationship
without the need for optical redshifts.

\subsection{ Expectations for ``spectral lags'' in GRBs }

Via analysing a set of 7 GRBs with directly determined redshifts
Norris, Marani and Bonnell (1999) suggested
that a temporal lag $\tau$ exists between neighbouring spectral
channels of BATSE GRB recordings.
Moreover they find that the isotropic luminosity L$_{53}$
in units of 10$^{53}$ ergs/sec depends on the lag $\tau$ as:
\begin{equation}
L_{53} \approx 1.3 \times (\tau/0.01 sec)^{k}
\label{lag}
\end{equation}
and find that k=$-$1.14 gives the best fit to the data.
If such a lag exists, the cannonball model firmly predicts
that k=$-$3 because I$_s$ $\sim$ L$_{53}$ and W $\sim$ $\tau$
in (eq.(\ref{iw}). This would seem to be evidence against the
cannonball model.
\\
However, using a somewhat different method to define the lag,
Wu and Fenimore (1999) find lag values for the same
bursts which are a factor 3 - 7 larger than the ones of Norris et al..
The reasons for this discrepancy 
are not clear, but are likely connected with the
fact that the lags are very small 
(milli-seconds) for the most luminous bursts.
Because the disagreement
turns out to be largest for the smallest lags one would obtain a smaller
value of the exponent in eq. (\ref{lag}) with the values
of Wu and Fenimore (2000), in better agreement with the expectation
of the cannonball model.
It is probably wise to wait
until the ``dust settles'' on the technically tricky issue
to extract very small lags via cross-correlation,
before drawing further quantitative conclusions.
\\
GRB 980425 with a
very small L$_{53}$ has a very large spectral lag $\tau$=4.5 seconds.
It was recently pointed out by Salmonson (2000) that this
value and the one for the GRB with the next longest lag
(GRB 970508 with $\tau$=0.4 seconds) does in fact lead to an
index of $-$3 in eq.(\ref{lag})(the one expected in the cannonball model).
Like me
Salmonson (2000) interprets the lag-luminosity 
relationship as a kinematical
effect in GRB-jets but his assumed geometry is
qualitatively different from the cannonball model
and he obtains different predictions for 
the exponent in eq.(\ref{lag}).

\section{Discussion}

It has been shown that the variability - luminosity relationship
empirically proposed by Fenimore and Ramirez-Ruiz (2000),
the lack of cosmological time dilation in GRB spikes
as a function of variability derived redshifts and the
functional dependence of spike intensities on their width
(\cite{feni2})
can be quantitatively understood in the cannonball model
(\cite{darrujula}) for GRBs.
This is evidence for the cannonball model
and for the reality of the relationship.
If the cannonball model is correct, GRBs are not very much
rarer than SNe. Independent of any model-details the combined
cannonballs have to  pierce through an overlying surface mass density
corresponding to a few solar masses 
while remaining ultra-relativistic. This is only possible
if their initial combined energy is $>$ 10$^{52}$ ergs.
With these boundary conditions (but {\it not} within standard jet
model (\cite{pugliese2}))
it is
virtually unavoidably that cannonballs
play a major role in the production
of Galactic cosmic rays at all energies
(\cite{dar98b,darplaga}).
If the variability - luminosity relationship is correct, it is bound
to play an important role in observational cosmology. 
\\
Cannonball-model predictions for the cosmological redshift in the GRB structure
at low frequencies and the precise form of the lag - luminosity
relationship have been made.
This might lead to further evidence for (or rejection of) the cannonball model
and improvements of the variability - luminosity relationship, which would be
crucial for its use in cosmology.
The road to ``smoking-gun evidence'' for the cannonball model
has been traced out by Dar and De R\'ujula
(2000a): 
search for superluminal motion in the remnant of GRB 980425/SN1998bw!

\begin{acknowledgements}
The major results of this paper were born in discussions with
Arnon Dar and Alvaro De R\'ujula about their cannonball model. 
I thank Alvaro De R\'ujula for hospitality at CERN and him, Arnon Dar,
Maria Diaz, J\"urgen Gebauer and Silvia Pezzoni for suggestions
on a draft for this manuscript. The author was supported by a Heisenberg
Fellowship of the DFG.
\end{acknowledgements}

\end{document}